\documentclass[12pt,noshowpacs,nofootinbib,notitlepage,amsmath]{revtex4-1}
\usepackage{setspace}
\allowdisplaybreaks
\usepackage{graphicx,color}
\usepackage[colorlinks=true,citecolor=blue,linkcolor=blue,urlcolor=blue]{hyperref}
\usepackage[charter]{mathdesign}
\usepackage{multirow}
\usepackage{enumerate}
\usepackage{slashed}

\DeclareSymbolFontAlphabet{\mathcal}{symbols}
\DeclareSymbolFont{symbols}{OMS}{xmdcmsy}{m}{n}
\DeclareSymbolFont{largesymbols}{OMX}{xmdcmex}{m}{n}
\SetSymbolFont{symbols}{bold}{OMS}{xmdcmsy}{b}{n}
\begin{document}  
\title{\color{blue}\Large A bound state model for a light scalar}
\author{Bob Holdom}
\email{bob.holdom@utoronto.ca}
\affiliation{Department of Physics, University of Toronto, Toronto, Ontario, Canada  M5S1A7}
\author{Roman Koniuk}
\email{koniuk@yorku.ca}
\affiliation{Department of Physics, York University, Toronto, Ontario, Canada  M3J1P3}
\begin{abstract}
Recent lattice studies of near-conformal strong dynamics suggest the existence of a light scalar. This provides motivation to consider a simple Hamiltonian-based bound-state model where the pseudoscalar, scalar, vector and axial-vector states are treated on an equal footing. The model interpolates between the non-relativistic limit and the highly relativistic chiral limit, where the pseudoscalar mass drops to zero. The fermion mass becomes purely dynamical at this point. When the gauge coupling is constant over a moderate range of scales the scalar becomes significantly lighter than the spin 1 states as the chiral limit is approached. We relate this result to the behavior of the form factors of the respective states and their decay constants. In the conformal limit of the model all masses vanish and the theory is characterized by scaling dimensions.
\end{abstract}
\maketitle

It has long been speculated that some near-conformal strong gauge dynamics underlies electroweak symmetry breaking. It has also been speculated that such dynamics could give rise to an unusually light scalar particle. However convincing theoretical arguments for such a state have proven to be elusive. Near-conformal dynamics naively suggests a light dilaton, but in gauge theories it is difficult to establish any limit in which chiral symmetry breaking and a massless dilaton can occur simultaneously. The discovery of the light Higgs boson has served to focus more attention on this issue.

In view of this it is especially interesting that lattice studies of near-conformal strong gauge theories report evidence of a scalar state that is significantly lighter than what occurs in QCD-like theories. Three groups \cite{Appelquist:2016viq,Aoki:2016wnc,Aoki:2014oha,Fodor:2016pls,Fodor:2012ty} find that an SU(3) gauge theory with a sufficient number of light fermions generates a light singlet scalar state with a mass similar to the mass of  the pseudoscalar state over the range of explicit fermion masses that have been studied. These groups also claim that this occurs in a phase with dynamical chiral symmetry breaking.

If such a state is to play the role of the Higgs boson of the standard model then, to a good approximation, it must enter the low energy description as a fluctuation around the vacuum expectation value (vev) of a scalar doublet. Since the dynamical fermion mass is the order parameter of the underlying theory, around which there are both scalar and pseudoscalar fluctuations, the suggestion is that the scalar and pseudoscalar states should have similar couplings to the heavy fermions. These couplings to fermions are described by form factors, or in the language of quark models, by wave functions. Similar scalar and pseudoscalar form factors will help ensure that when the heavy fermions are integrated out, a standard light Higgs boson description emerges. Similar form factors may not be something to be expected of a dilaton interpretation. The chiral transformation and the scale transformation of the momentum dependent dynamical fermion mass are quite different, resulting in form factors of different shapes.\footnote{A zero momentum pion has a form factor proportional to the dynamical mass function $\Sigma(p)$, while for the dilaton it is $\Sigma(p)-p\Sigma'(p)$.}

Here we would like to shed some light on the following question: What is it about the near-conformal gauge dynamics that could lead to a near parity-doubled scalar-pseudoscalar sector, both for the masses and the form factors? A similar question pertains to the vector and axial-vector mesons, since the lattice studies indicate that the splitting between their masses, and their decay constants, is smaller than in QCD. We shall present a simple model of fermion bound states that allows us to study the relation between the mass spectrum of these four states and the behavior of their respective form factors as the chiral limit is approached.

The basic dynamics can be illustrated by an abelian gauge interaction. Our bound state model was developed in \cite{Dykshoorn:1990bp,r1} and it is based on the QED Hamiltonion in Coulomb gauge,
\begin{align}
H&= \int d^3x\,\left(\psi^{\dag}\{\boldsymbol\alpha\cdot\left[(1/i)\boldsymbol\nabla-e\mathbf{A}\right]+\beta m\}\psi + \frac{1}{2}(\mathbf{E}^2+\mathbf{B}^2)\right),\\
&\mathbf{E}^2 = \mathbf{E_t}^2 + \mathbf{E_l}^2 \ \ \  \mathbf{E_t} = -\mathbf{\dot {A}}, \ \ \ \mathbf{B} = \boldsymbol\nabla\times\mathbf{E},\nonumber\\
&\mathbf{E_l}^2 = \frac{e^2}{4\pi}\int d^3y\,\frac{\psi^{\dag}(\mathbf{x})\psi(\mathbf{x})\psi^{\dag}(\mathbf{y})\psi(\mathbf{y})}{|\mathbf{x}-\mathbf{y}|}.\nonumber
\end{align}
The two parameters of the model are the explicit fermion mass $m$ and the coupling $\alpha=e^2/4\pi$. The concept of a dynamical fermion mass will emerge.
 
\begin{figure}[h]
  \centering%
{ \includegraphics[width=14cm]{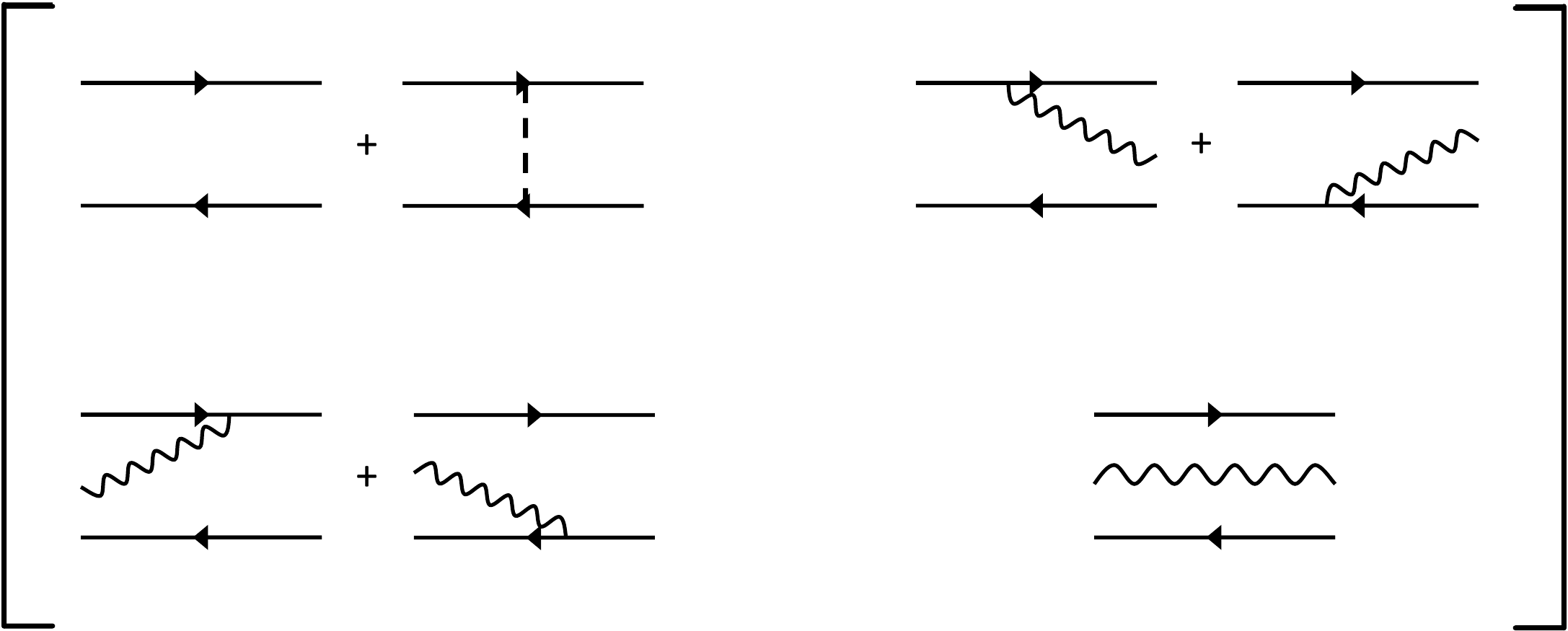}}
\caption{\label{fig1} 
The form of the Hamiltonian in the truncated Fock-space. The instantaneous Coulomb exchange is shown by the dashed line. Each diagram has one insertion of the Hamiltonian, with the sum over energy insertions in the top-left and bottom-right diagrams not shown explicitly.}
\end{figure}

We define a truncated Fock-space and then simply diagonalize the Hamiltonian in that space to determine the bound state spectrum. This will implement a type of ladder graph summation. In particular we consider the eigenvalue problem
\begin{align}
H\left[\begin{array}{c}|q \bar{q}\rangle \\|q \bar{q}\gamma\rangle\end{array}\right]=M\left[\begin{array}{c}|q \bar{q}\rangle \\|q \bar{q}\gamma\rangle\end{array}\right],
\label{e1}\end{align}
where the Hamiltonian in this space is shown schematically in Fig.~\ref{fig1}. We define
\begin{align}
|q\bar{q}\rangle = \sum_{ss'} \int \,d^3p\, F(\mathbf{p},s,s')|\mathbf{p},s;\mathbf{-p},s'\rangle
,\end{align}
where $F(\mathbf{p},s,s')$ contains the structure $f(p) \bar{u}(\mathbf{p},s)\Gamma v(\mathbf{-p},s')$.\footnote{We shall be more precise with the normalization factors when we discuss the decay constants.} The Dirac matrix $\Gamma$ is chosen to be $\gamma_5, \mathbf{1}, \boldsymbol{\gamma}, \gamma_5\boldsymbol{\gamma}$, for a pseudoscalar, scalar, vector, or pseudo-vector state, respectively. Similarly
\begin{align}
|\bar{q}q \gamma\rangle = \sum_{ss'\lambda} \int \,d^3p\, d^3q\,G(\mathbf{p},\mathbf{q},s,s',\lambda)|\mathbf{p},s;\mathbf{-p},s';-\mathbf{p}-\mathbf{q},\lambda\rangle,
\end{align}
where $G(\mathbf{p},\mathbf{q},s,s',\lambda)$ contains the structure $ \epsilon^\mu(-\mathbf{p}-\mathbf{q},\lambda)\bar{u}(\mathbf{p},s)\, \gamma_\mu\Gamma\, v(\mathbf{q},s')\,g(\mathbf{p},\mathbf{q})$.

The matrix equation (\ref{e1}) then generates two integral equations that are to be used to determine the functions $F$ and $G$ and the eigenvalue $M$. These equations can be rewritten as an integral equation for $F$ alone and a second equation that determines $G$ in terms of $F$. We are interested in the former, which takes the form of an integral equation for the form factor $f(p)$,
\begin{align}
M{f}(p) = 2\sqrt{p^2+m_{\rm tot}^2}\;{f}(p) -\frac{\alpha}{4\pi}\int \frac{q}{p}K(p,q;m){f}(q)\, dq.
\label{e4}\end{align}
The kernels $K(p,q;m)$ for the four channels (pseudoscalar $P$ ($0^{-+}$), scalar $S$ ($0^{++}$), vector $V$ ($1^{--}$), axial-vector $A$ ($1^{++}$)) are given in Appendix \ref{appA}. Through the diagonalization, the kernels incorporate both the instantaneous Coulomb interaction and real photon exchange. Solving the equation for the form factor $f(p)$ effectively sums the corresponding ladder graphs. Finding the solution with the lowest eigenvalue $M$ gives the mass of the lowest lying state in each of the channels.

The $m$ in the kernel $K(p,q;m)$ is the explicit fermion mass appearing in the Hamiltonian. The appearance of a total fermion mass $m_{\rm tot}$, in the fermion energy term in (\ref{e4}), occurs as follows. In addition to the integral term in (\ref{e4}), the diagonalization also generates fermion self-energy corrections. We don't keep these corrections explicitly, but we instead model their effect by modifying the mass in the fermion energy term. Thus $m_{\rm tot}$ includes a perturbative correction to $m$. The integral equation solution also supports a dynamical component in $m_{\rm tot}$, a dynamical mass that survives even when $m$ vanishes. We shall return to this below.

In the nonrelativistic limit our kernels reproduce the known corrections to the lowest lying bound state masses to order $\alpha^4$, as listed in Appendix \ref{appA}.  We have not shown additional diagrams involving the instantaneous Coulomb exchange in the bottom right entry in Fig.~\ref{fig1}. These diagrams generate corrections beyond $\alpha^4$ for masses after diagonalization and we have ignored these diagrams for simplicity. A nonabelian theory can be accommodated at our level of approximation by replacing $\alpha$ by $C_F\alpha$ where $C_F = (N_c^2-1)/2N_c$ for $SU(N_c)$.

\begin{figure}[h]
  \centering%
{ \includegraphics[width=12cm]{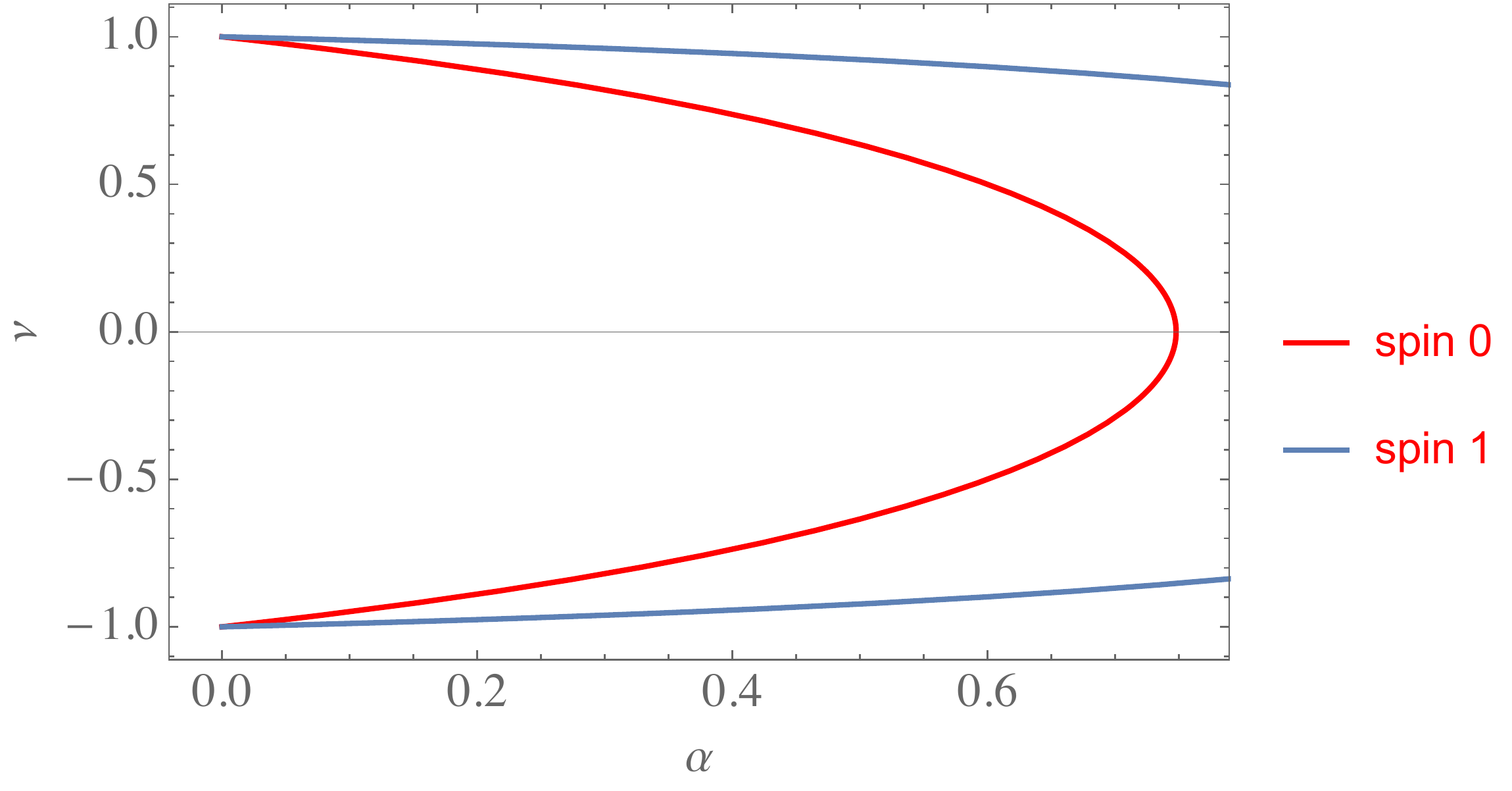}}
\caption{\label{fig1a} 
The scaling dimension $\nu(\alpha)$ in the spin 0 and 1 form factors when $m=0$.}
\end{figure}

We first consider the conformal limit of the model where $m=m_{\rm tot}=M=0$. In this limit the dynamics displays a parity doubling, since $K(p,q;0)$ is degenerate for the spin 0 states and the spin 1 states respectively.\footnote{Note that this is unlike the degeneracy of the non-relativistic limit where the pseudoscalar/vector and scalar/axial-vector pairs are degenerate.} In each sector and for a given coupling the integral equation possesses a power law solution $f(p)=1/p^{2+\nu(\alpha)}$. The scaling dimensions $\nu(\alpha)$ for the two sectors are displayed in Fig.~\ref{fig1a}, and in the spin 0 case $\nu(\alpha)$ is given implicitly by
\begin{align}
\alpha=\frac{4\nu(\nu^2-1)\cot(\pi\nu/2)}{3-3\nu^2+\nu^2\cot(\pi\nu/2)^2}.
\end{align}
At $\nu=0$, $\alpha$ reaches its maximum value $\alpha_{\rm max}=8\pi/(4+3\pi^2)\approx0.748$, as also obtained in a Bethe-Salpeter approach \cite{Sucher}.

The power law behavior in the conformal limit will tell us about the asymptotic power law behavior when $m>0$. In particular, the form factor of the lowest lying state in each sector for a given $\alpha$ will have an asymptotic scaling dimension given by the upper branch, $\nu(\alpha)>0$.\footnote{The behavior of the two branches are reminiscent of the regular and irregular solutions in a Schwinger-Dyson analysis, but in our context the negative branch is not relevant to the lowest lying state.} That is, as the form factors evolve away from the nonrelativistic wave functions at small coupling towards larger coupling, they fall less and less quickly in the UV. Since $\nu(\alpha)$ decreases more in the spin 0 sector, for increasing $\alpha$, we expect that the spin 0 form factors become more heavily weighted in the UV relative to the spin 1 form factors. 

\section*{Chiral symmetry breaking}
As we move away from the conformal limit and turn on $m$ and $m_{\rm tot}$, the quantity $m_{\rm tot}$ is more relevant to the mass spectrum. In particular at weak coupling the explicit mass $m$ approaches $m_{\rm tot}$, the spin 1 masses are just below $2m_{\rm tot}$, and the result is a description of the heavy quark bound states of QCD. Of more interest here is strong coupling where a chiral limit can be approached with $m$ tending to zero. In this limit $m_{\rm tot}$ is purely dynamical, appropriate for describing the light quark bound states of QCD. We use a vanishing pseudoscalar mass to define the chiral $m\to0$ limit, and we approach this limit by raising $\alpha$ while holding $m_{\rm tot}$ fixed. The explicit mass $m$ must range from $m_{\rm tot}$ to 0, as $\alpha$ ranges from 0 to $\alpha_c$, and in particular we expect $m\sim m_\pi^2$ at small $m$. Thus the explicit mass is an {\it implicit} function of $\alpha$, $m(\alpha)$. It is in this way that we model the approach to the chiral limit for near-conformal chiral symmetry breaking dynamics. At vanishing pseudoscalar mass, we recover a gap equation that determines the critical coupling needed to generate a purely dynamical fermion mass. In the following, the approach to the critical coupling $\alpha_c$ will be synonymous with the approach to the chiral limit.

As it now stands the integral equation still depends on the not completely determined implicit function $m(\alpha)$. This integral equation also has the following undesirable property in the chiral limit, when $m(\alpha_c)=0$ and $m_{\rm tot}\neq0$. The remaining mass $m_{\rm tot}$ in the integral equation does not appear in the kernel $K(p,q;m(\alpha))$. The implication is that the spectrum is parity doubled in the chiral limit, just as it was in the conformal limit. This property is an artifact of the order in perturbation theory that we have used to derive the integral equation. At higher order, self energy corrections will appear \textrm{inside} the kernel of the integral equation and thus, according to our prescription, dynamical mass contributions are also introduced there. This will remove the parity doubling; the scalar mass will no longer vanish in the chiral limit, which is still defined by where the pseudoscalar mass vanishes. In this way we are led to consider a better specified and physically motivated modification of our integral equation where we replace the kernel $K(p,q;m(\alpha))$ by the kernel $K(p,q;m_{\rm tot})$. We shall compare the results of this new integral equation to another one where the kernel is replaced by $K(p,q;0)$. The latter gives a parity doubled spectrum for all $\alpha$. A fundamental surprise of our work will be the unexpected similarities of the results of these two integral equations.

The details of our numerical procedure are given in Appendix \ref{appB}. We find that the integral equation is well behaved except for the extremely slow convergence \textrm{very} close to the chiral limit. We shall impose a UV cutoff, which not only alleviates the numerical problem, but which also better represents the physical situation of interest. Namely, if a near-conformal strong dynamics underlies electroweak symmetry breaking, it likely only exists over a finite range of energy scales. In our results we shall consider two cases, a physically interesting cutoff of a moderate size, $\Lambda/m_{\rm tot}=\sinh(8)\approx1490$, and a very high cutoff, $\Lambda/m_{\rm tot}=\sinh(34)\approx2.9\times 10^{14}$. The corresponding critical couplings for the chiral limit are $\alpha_c=0.816$ and 0.752. From this we see that $\alpha_c$ approaches the $\alpha_{\rm max}$ of the conformal limit, from above, in the infinite cutoff limit. 

\begin{figure}[h]
  \centering%
{ \includegraphics[width=13cm]{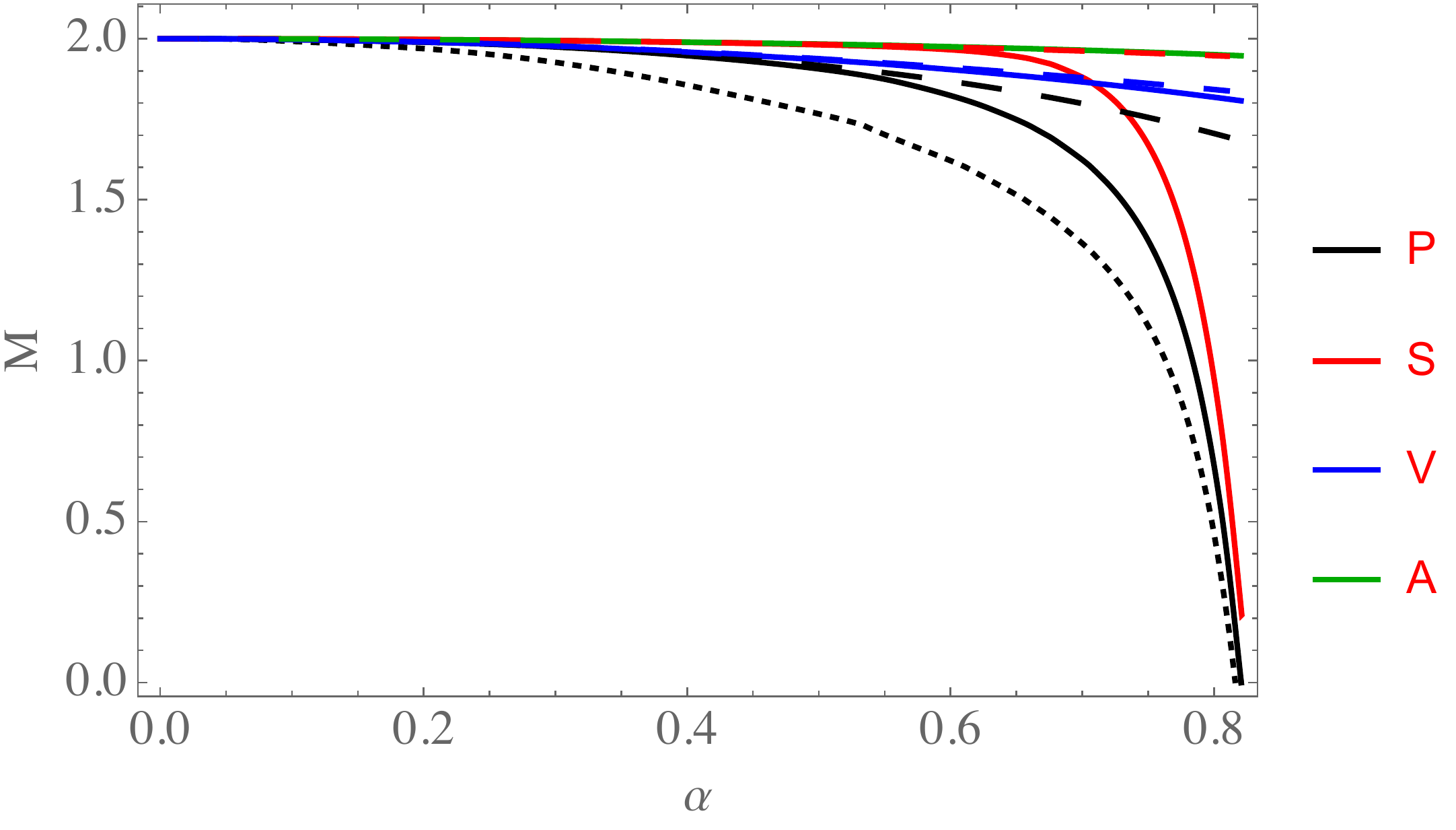}}
\caption{\label{fig2} 
The pseudoscalar, scalar, vector and axial-vector masses as a function of the gauge coupling with the kernel $K(p,q;m_{\rm tot})$ and $m_{\rm tot}=1$. The moderate UV cutoff is used. The dashed lines are the perturbative results to order $\alpha^4$. The dotted line is the spin 0 mass result with the kernel $K(p,q;0)$.}
\end{figure}


In Fig.~\ref{fig2} we show the four masses as a function of the coupling for the kernel $K(p,q;m_{\rm tot})$. We also display the perturbative results to order $\alpha^4$ that are given in Appendix \ref{appA}. From the deviations from the perturbative curves we see that both the pseudoscalar and scalar states are becoming highly relativistic close to the critical coupling. Since the form factors of these states are also becoming more UV dominated we understand why the relative impact of $m_{\rm tot}$ on the physical masses is reduced. The vector and axial-vector masses remain relatively close to the weak coupling limit value of $2m_{\rm tot}$ with only a small relative mass splitting. For the kernel $K(p,q;0)$ that gives the patity doubled spectrum, the resulting spin 0 mass curve is also shown in Fig.~\ref{fig2}.

\begin{figure}[h]
  \centering%
{ \includegraphics[width=13cm]{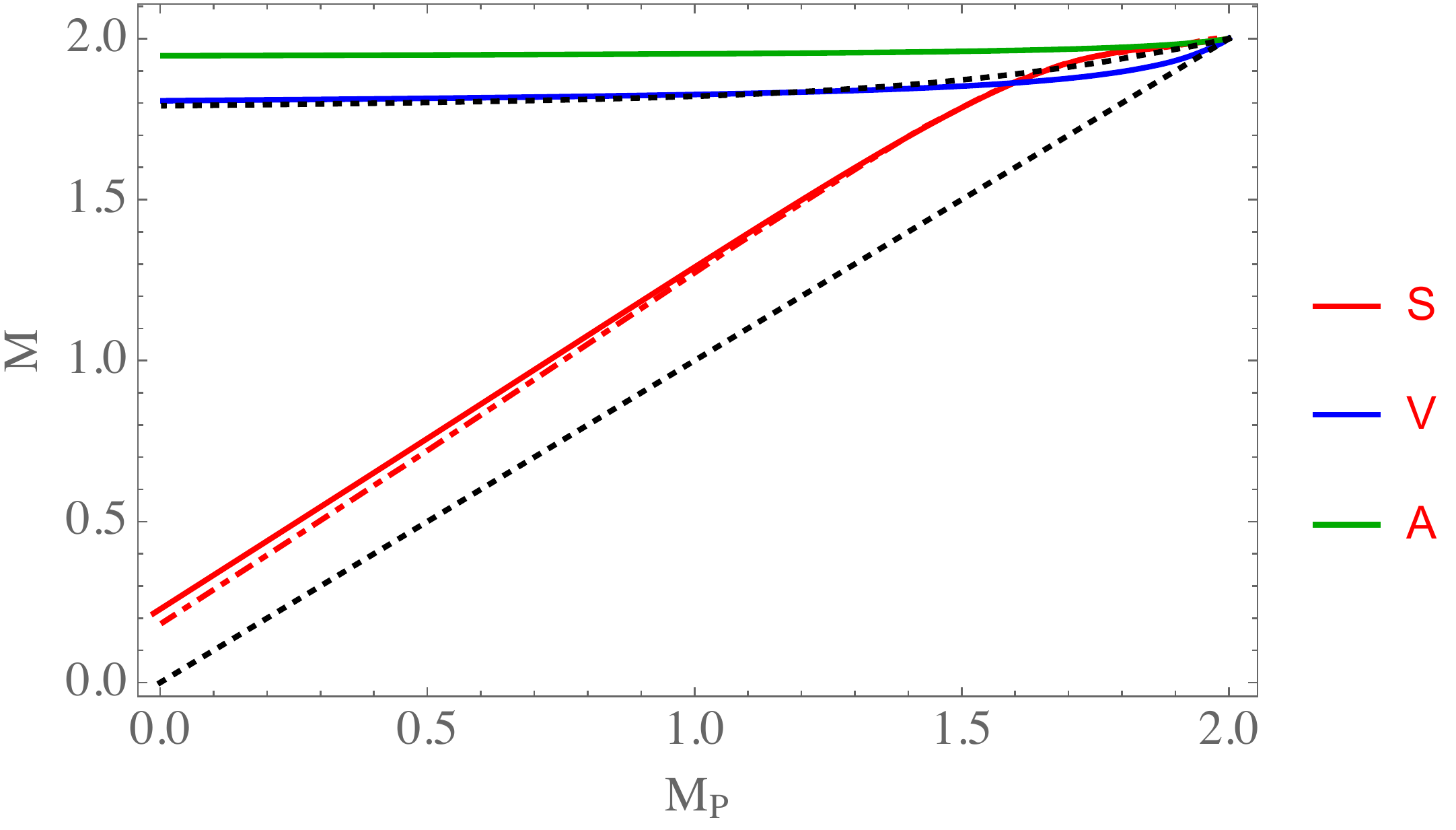}}
\caption{\label{fig3} 
The scalar, vector and axial-vector masses as a function of the pseudoscalar mass with the kernel $K(p,q;m_{\rm tot})$ and $m_{\rm tot}=1$. Changing to the very high cutoff gives the dot-dashed line. The upper and lower dotted lines are are the spin 1 and spin 0 results when the kernel $K(p,q;0)$ is used.}
\end{figure}

The relation among the four masses can be displayed differently by plotting three masses as a function of the pseudoscalar (the fourth) mass. This is shown in Fig.~\ref{fig3} for the kernel $K(p,q;m_{\rm tot})$. In this form it can be compared more directly with lattice results where masses are given as functions of the explicit fermion mass. The scalar mass becomes small as the pseudoscalar mass vanishes. Note that it shows only a slight further decrease when we use our very high cutoff. For the parity doubled kernel $K(p,q;0)$, the scalar mass curve is just the straight line $M_S=M_P$. Our primary result is that even though the kernel $K(p,q;m_{\rm tot})$ destroys the parity doubling, the resulting scalar mass still becomes small in the chiral limit.

\begin{figure}[h]
  \centering%
{ \includegraphics[width=13cm]{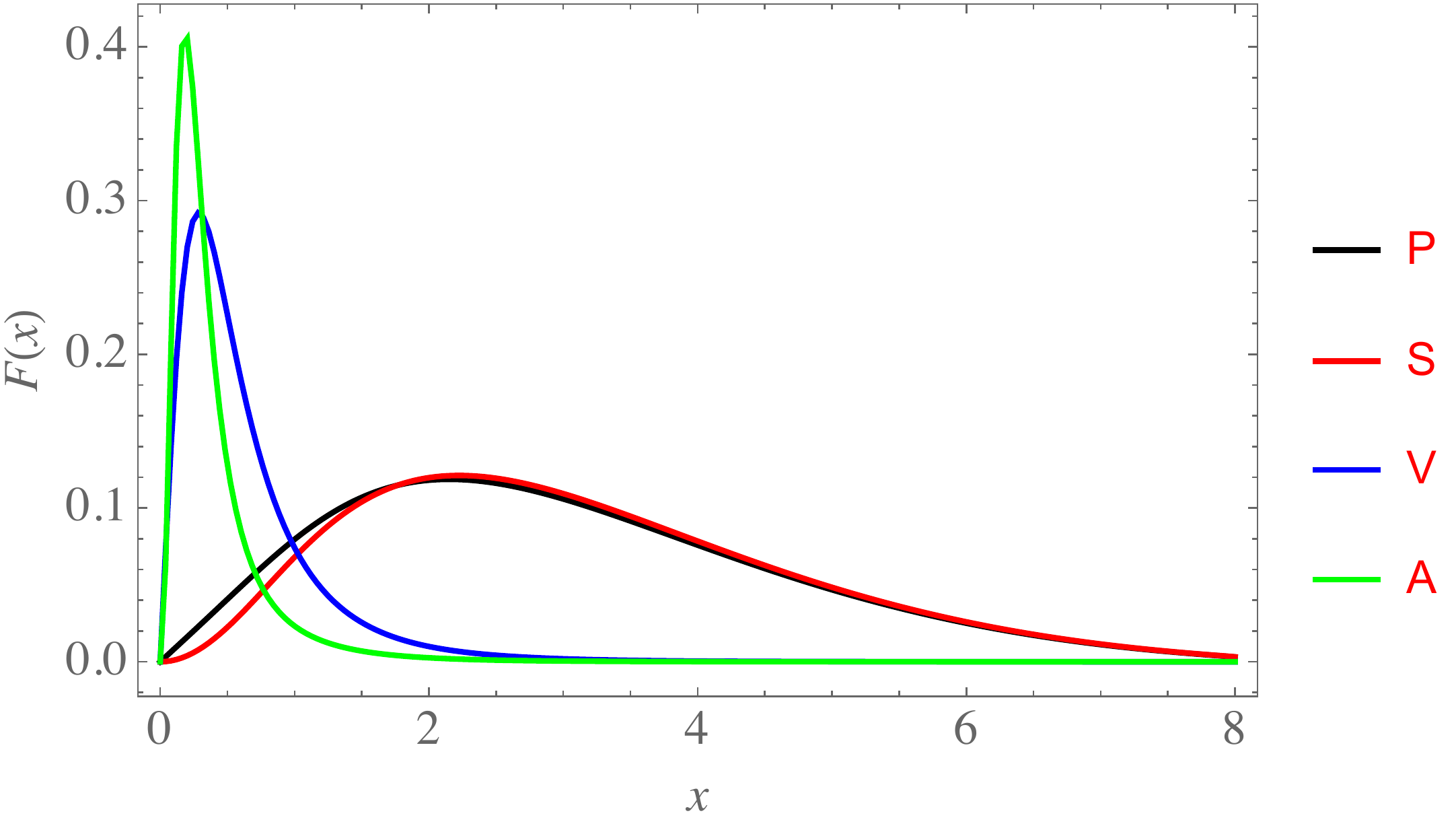}}
\caption{\label{fig5} 
Form factors for moderate UV cutoff at the critical coupling. $F(x)= p\,(p^2+1)^\frac{1}{4}{f}(p)$ with $p=\sinh(x)$ and $m_{\rm tot}=1$.}
\end{figure}

Better understanding of the mass spectra for the kernel $K(p,q;m_{\rm tot})$ can be obtained by inspecting the form factors at the critical coupling in Fig.~\ref{fig5}. $F(x)$ is as defined in Appendix \ref{appB}. We see the clear separation between the spin 0 and spin 1 form factors, with the highly relativistic nature of the former evident by the much larger typical momenta. The near parity doubled nature of the spin 0 sector is apparent in these form factors, where the effect of $m_{\rm tot}$ in the kernel is visible only in their difference at low momentum. When the very high cutoff is considered instead, there is little further change of the spin 0 form factors.

\begin{figure}[h]
  \centering%
{ \includegraphics[width=13cm]{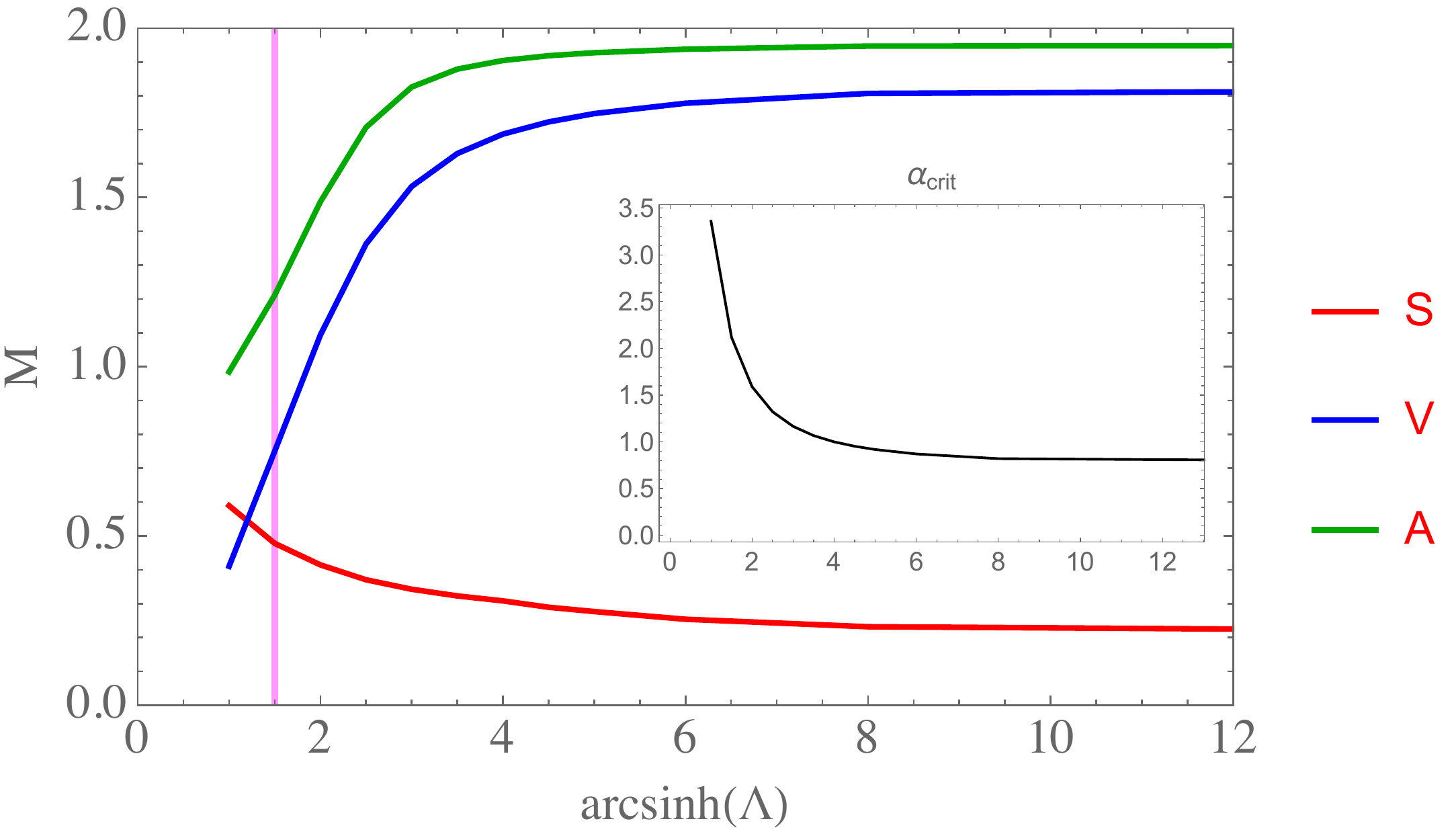}}
\caption{\label{fig4} 
Masses at the critical point as a function of the UV cutoff $\Lambda$ with $m_{\rm tot}=1$. Insert shows $\alpha_{\rm crit}(\Lambda)$.}
\end{figure}

\begin{figure}[h]
  \centering%
{ \includegraphics[width=13cm]{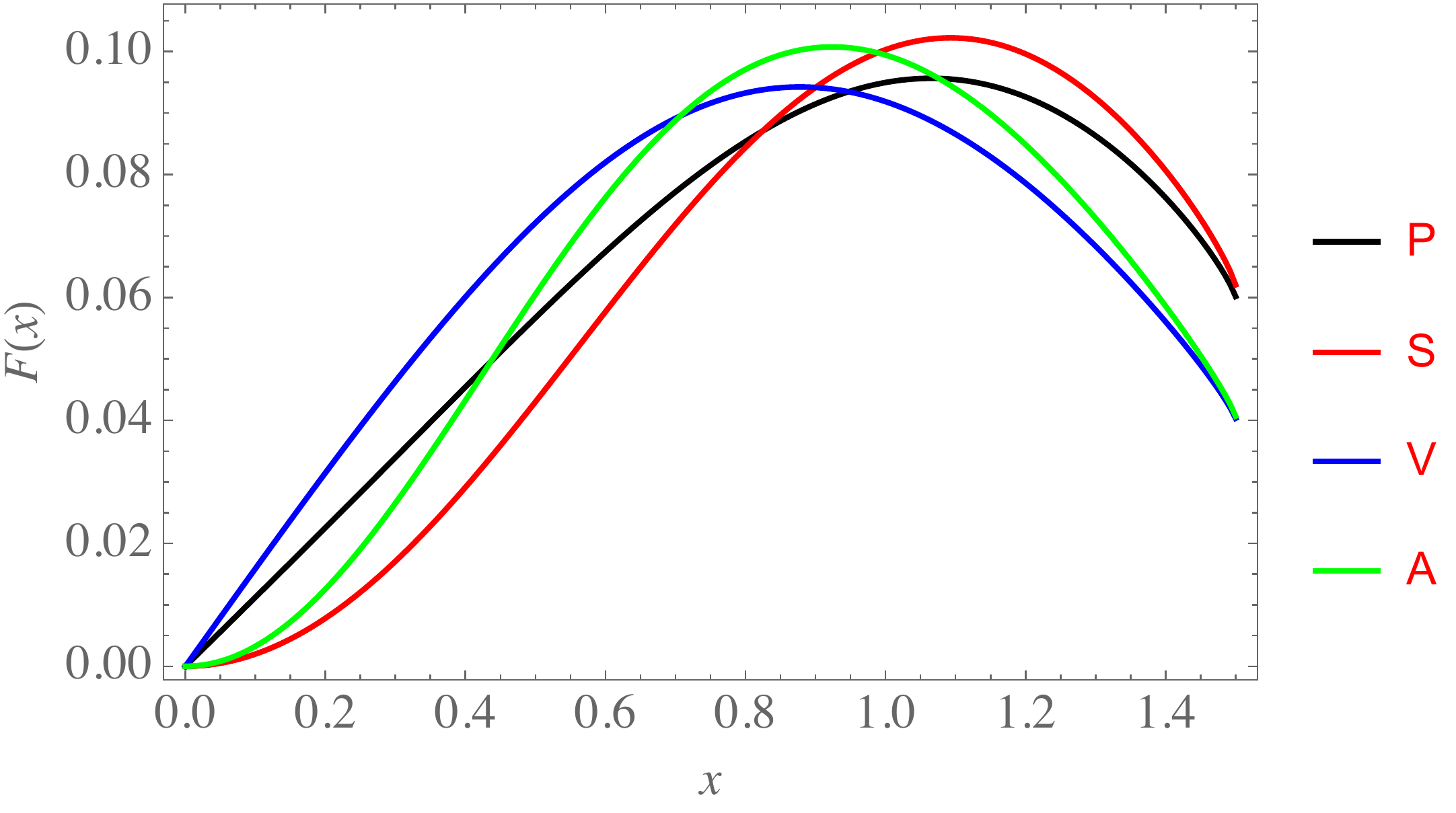}}
\caption{\label{fig6} 
Form factors for low UV cutoff at the critical coupling.}
\end{figure}

Our results indicate that a light scalar is related to the presence of conformal dynamics (a constant coupling), over a moderate range of momentum scales. To show this more explicitly we study the effect of lowering the cutoff. Fig.~\ref{fig4} shows the mass spectrum in the chiral limit as a function of the cutoff. As the cutoff decreases, not only does the scalar mass increase, but in addition the spin 1 masses drop while their mass splitting increases. In fact for a cutoff around $\Lambda/m\approx \sinh(1.5)\approx2.1$ the spectrum becomes ``QCD-like''. This may be expected since a constant coupling in a theory with a small cutoff roughly models QCD where the coupling is strong for a small range of momenta before falling quickly. The insert in Fig.~\ref{fig4} shows that the value of the coupling must also increase for decreasing cutoff. Fig.~\ref{fig6} shows the four form factors for the QCD-like spectrum. The spin 0 and spin 1 form factors are now much more similar, while at the same time the effect of $m_{\rm tot}$ is more pronounced. This low cutoff model may not be representative of real QCD, since all four states are quite highly relativistic and strongly bound, and the effective $m_{\rm tot}$ would have to be relatively large.

\section*{Decay constants}
\begin{figure}[h]
  \centering%
{ \includegraphics[width=12cm]{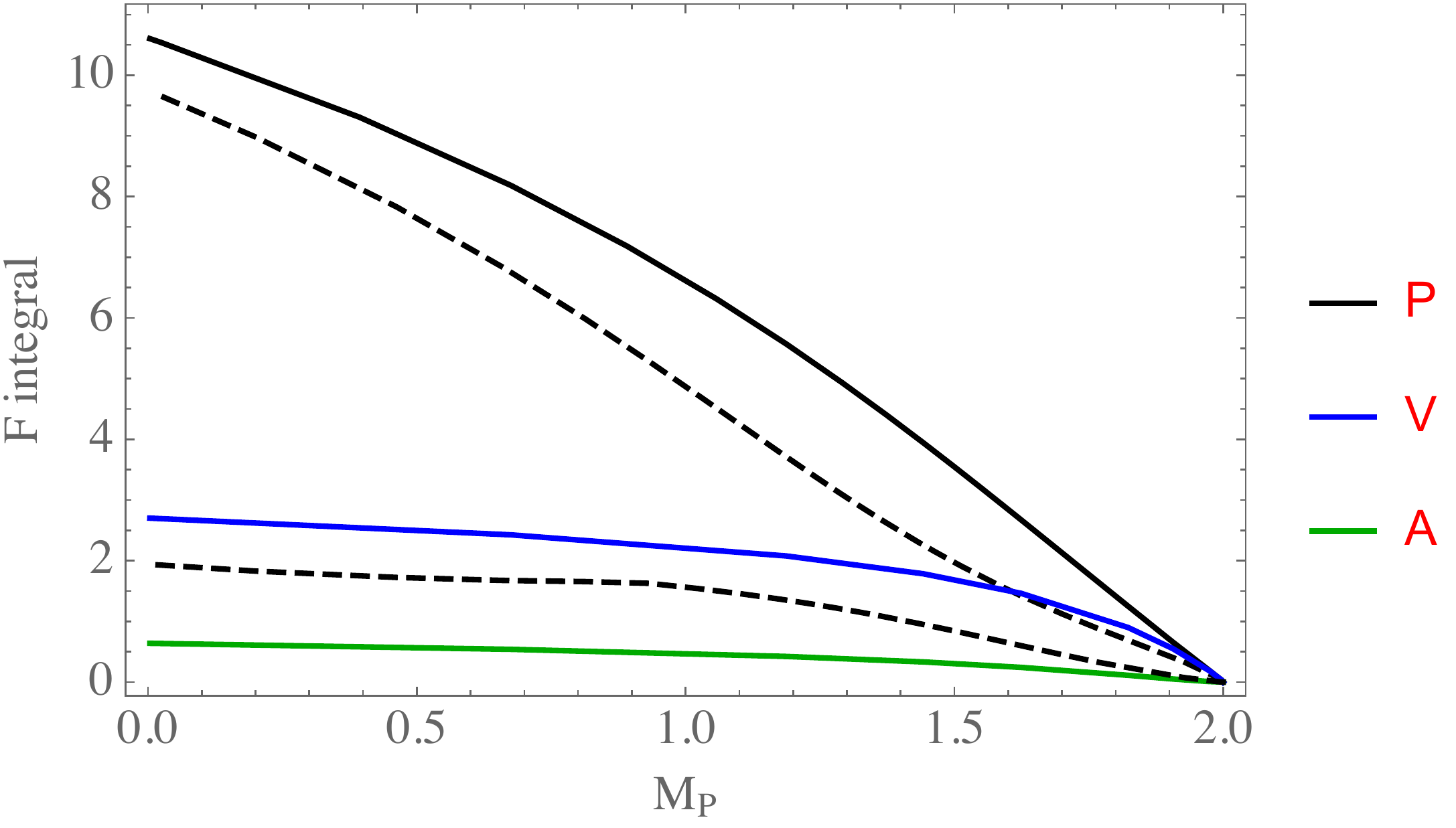}}
\caption{\label{fig7} 
The integrals appearing in equations (\ref{e5}, \ref{e7}, \ref{e8}) for the $P$, $V$ and $A$ decay constants with $m_{\rm tot}=1$. The upper and lower dotted lines are the spin 0 and spin 1 results when the kernel $K(p,q;0)$ is used while also setting $m=0$ in the integrands.}
\end{figure}

In QCD $f_{\pi}$ is defined through the coupling of the axial vector current to the pion,
\begin{align}
\langle 0 | A_{\mu}(x) | \pi^-(P)\rangle = \sqrt{2}P_{\mu}f_{\pi}e^{iP\cdot x},
\label{e6}\end{align}
corresponding to $f_\pi\approx93$ MeV. We can obtain analogous quantities in our one flavor model for the states $| q\bar{q}\rangle=| P\rangle, |V\rangle$ and $|A\rangle$. Our states are defined to have a rest frame normalization $\langle q\bar{q} | q\bar{q}\rangle = 2M\delta^{(3)}(0)$, so that
\begin{align}
|q\bar{q}\rangle =\sqrt{2M} \int \frac{d^3p\,\phi(\mathbf{p})}{(2\pi)^{3}2E_p}\,\sum_{ss'}\sum_{ij}\frac{\delta_{ij}}{\sqrt{N_c}}\,\chi_{ss'}(\mathbf{p})\,|\mathbf{p},s;\mathbf{-p},s'\rangle
.\end{align}
We keep color factors and we have introduced $\phi(\mathbf{p})$ and  $\chi_{ss'}(\mathbf{p})$ such that
\begin{align}
&\int\phi(\mathbf{p})^2 \frac{d^3p}{(2\pi)^3}=\int f(p)^2p^2dp=1=\sum_{ss'}|\chi_{ss'}(\mathbf{p})|^2,\nonumber\\
&\chi_{ss'}(\mathbf{p}) = N(p)\bar{u}(\mathbf{p,s})\Gamma v(\mathbf{-p,s'}),\quad N(p)^{-2} ={ \rm{Tr}}[(\slashed{p}+m)\Gamma(\slashed{p}^{\dag}-m)\Gamma].\nonumber
\end{align}
Thus the definition of a decay constant as in (\ref{e6}) must be used in the rest frame and therefore for the three states
\begin{align}
&\langle 0 | A_0(x) | P\rangle = \sqrt{2}M_P F_P,\\
&\langle 0 | V_i(x) | V\rangle = \sqrt{2}M_V F_V\epsilon_i,\\
&\langle 0 | A_i(x) | A\rangle = \sqrt{2}M_A F_A\epsilon_i.
\end{align}
We have
\begin{align}
\sqrt{2}M F &=  \langle 0|\bar{\psi}(0)\tilde{\Gamma}\psi(0)|q\bar{q}\rangle\\
&=\sqrt{2M N_c}\int \frac{d^3p}{(2\pi)^3}\frac{1}{2E_p}N(p){ \rm{Tr}}[(\slashed{p}+m)\Gamma(\slashed{p}^{\dag}-m)\tilde{\Gamma}]\,\phi({\bf p})
,\end{align}
where $\Gamma = \gamma^5, \tilde{\Gamma} = \gamma^0\gamma^5$ for $P$, $\Gamma = \gamma^j,\tilde{\Gamma} = \gamma^i$ for $V$ and $\Gamma = \gamma^j\gamma^5, \ \tilde{\Gamma} = \gamma^i\gamma^5$ for $A$.
We end up with\footnote{We obtain the standard non-relativistic result for $F_P$ \cite{bernard} if we replace $E_p$ by $m$ and note that $\int\frac{d^3p}{(2\pi)^{3}}\phi(\mathbf{p})=|\psi(\mathbf{0})|$ where $\psi$ is the Fourier transform of $\phi$.}
\begin{align}
F_P\sqrt{M_P} &=\frac{\sqrt{N_c}}{\pi}\;m\int \frac{f(p)}{E_p}\,p^2dp\label{e5},\\
F_V\sqrt{M_V} &=\frac{\sqrt{N_c}}{\pi}\int \frac{f(p)}{E_p}\sqrt{\frac{2}{3}p^2+m^2}\;p^2dp\label{e7},\\
F_A\sqrt{M_A} &=\frac{\sqrt{N_c}}{\pi}\int \frac{f(p)}{E_p}\sqrt{\frac{2}{3}}\,p^3dp.\label{e8}
\end{align}
Fig.~\ref{fig7} shows the three integrals in these expressions. The $M_P$ dependence enters through $f(p)$, where as before we have traded $\alpha$ dependence for $M_P$ dependence. Results for the two choices of the kernel can be compared.

For the pseudoscalar at small $m$, close to the chiral limit,  (\ref{e5}) looks quite unlike chiral perturbation theory. On the other hand the lattice data also displays an odd behavior $F_P^2 \sim m$ \cite{Appelquist:2017wcg} in addition to the expected $M_P^2 \sim m$. In fact with $M_P^2 \sim m$, then the integral in (\ref{e5}) is proportional to $1-\sqrt{m/\kappa}$ for small $m$ and some mass scale $\kappa$. Then we find that (\ref{e5}) gives a close to linear behavior $F_P^2 \sim m$ in the range $0\lesssim m\lesssim 0.2\kappa$. For larger $m$, $F_P$ must eventually decrease and vanish at $m=m_{\rm tot}$ (vanishing $\alpha$). It should be noted again that we are able to reach the weak coupling limit by holding $m_{\rm tot}$ fixed while varying $\alpha$, and with the explicit mass $m$ varying accordingly. These are not the conditions under which the explicit mass is varied in the lattice simulations.

\section*{Conclusion}
Our model is able to relate the existence of a light scalar that is well separated from the heavier states, to the existence of near conformal dynamics that extends over a relatively wide range of scales. Thus the bound-state dynamics of a simple abelian gauge theory may be capturing some of the essential features of lattice studies of near conformal strong gauge dynamics. Our main observation is that the small scalar mass is related to scalar and pseudoscalar form factors that are heavily skewed towards the UV, where they become very similar, and where they become quite insensitive to the presence of the dynamical fermion mass. Thus masses and form factors of the spin-0 sector can be described as being close to the parity doubled limit. As we have mentioned in the introduction, this means that the light scalar has characteristics that differ from those expected of a light dilaton. The basic features of the form factors we have described for near conformal dynamics could be explicitly checked by a lattice measurement of closely related form factors (see for example \cite{Hietanen:2013fya}).

Are these conclusions related to some shortcomings of our model? As with a Schwinger-Dyson (SD) approach, the kernel of our bound-state integral equation is only calculated to a low order in perturbation theory. And although the physical bound state masses and the total quark mass is accessible, the current quark mass is not, which is quite unlike the SD approach. The form factors of our states are also accessible, but interactions among the states are not so easily obtained, and the connection to some effective action description is obscure. On the other hand we can compare our form factors to those obtained from nonlocal effective actions, with the latter constructed such that the stationarity condition gives the SD equation. There the pion is a chiral fluctuation around the order parameter. Similarly the dilaton is identified as a fluctuation that corresponds to a scale transformation. The pion and dilaton form factors that are determined in this way have quite different momentum dependence, as we have already indicated. But it turns out that no light dilaton emerges in such an approach  \cite{Holdom:1986ub,Holdom:1987yu}.  Thus our finding of a light scalar emerging without dilatonic properties, with a form factor closely resembling that of the pion, is consistent with this. Light dilatons do appear in approaches that are more local, such as where there are elementary scalar fields with Yukawa couplings and local potentials with perturbative corrections \cite{Grinstein:2011dq,Antipin:2011aa}. These local theories that find dilatons are very different from the non-local effective theories that do not find them. We do not see any significant inconsistency between these very different frameworks.

\appendix\section{Kernels}\label{appA}
The kernels for the four channels, with $E_p\equiv\sqrt{p^2+m^2}$, are the following \cite{Dykshoorn:1990bp,r1}. The pseudoscalar kernel was also obtained in a Bethe-Salpeter motivated approach \cite{Sucher}.
\begin{align}
&K(p,q;m)_{P} =\frac{E_p^2+E_q^2+6E_pE_q-4m^2}{E_pE_q}\log\left|\frac{p+q}{p-q}\right| -\frac{2pq}{E_pE_q}\\
&K(p,q;m)_{S}=\frac{(E_p^2+E_q^2)(E_pE_q-m^2)+6E_pE_q(E_pE_q+m^2)-4m^2(E_p+E_q)^2+4m^2}{pqE_pE_q}\log\left|\frac{p+q}{p-q}\right|\nonumber\\
&\ \ \ \ \ \ \ \ \ \ \ \ \ \ \ \ \ \ \ \ -2-\frac{2m^2}{E_pE_q}\\
&K(p,q;m)_{V} = \frac{4(E_p^2+E_q^2)[(E_p+E_q)^2-4m^2]-16m^2(E_pE_q-m^2)}{4pqE_pE_q}\log\left|\frac{p+q}{p-q}\right]\nonumber\\
&\ \ \ \ \ \ \ \ \ \ \ \ \ \ \ \ \ \ \ \ -4 - \frac{2E_p}{E_q}- \frac{2E_q}{E_p}+\frac{4m^2}{E_pE_q}\\
&K(p,q;m)_{A} =\frac{2pq}{[(2E_p^2+m^2)(2E_q^2+m^2)]^{1/2}}\nonumber\\
&\ \ \ \ \ \ \ \ \ \ \ \ \ \  \times\Bigg[ \frac{4(E_p^2+E_q^2)[(E_p+E_q)^2+2m^2]-2m^2(E_p+E_q)^2-8m^2(E_pE_q+m^2)}{4pqE_pE_q}\log\left|\frac{p+q}{p-q}\right|\nonumber\\
&\ \ \ \ \ \ \ \ \ \ \ \ \ \ \ \ \ \ \ \ -4 - \frac{2E_p}{E_q}- \frac{2E_q}{E_p}-\frac{3m^2}{E_pE_q}\Bigg]\
\end{align}
\noindent These kernels reproduce the following perturbative corrections that were first calculated for positronium in \cite{ferell}.
\begin{eqnarray}
P :\ \ \  \frac{M}{m} &=& 2-\frac{\alpha^2}{4}-\frac{63\alpha^4}{192}\\
S :\ \ \  \frac{M}{m} &=& 2-\frac{\alpha^2}{16}-\frac{95\alpha^4}{3072}\\
V :\ \ \  \frac{M}{m} &=& 2-\frac{\alpha^2}{4}+\frac{\alpha^4}{192}\\
A :\ \ \  \frac{M}{m} &=& 2-\frac{\alpha^2}{16}-\frac{47\alpha^4}{3072}
\end{eqnarray}

\section{Numerics}\label{appB}
We write the integral equation in (\ref{e4}) as 
\begin{align}
M\frac{{g}(p)}{\bar{E}_p} = 2{g}(p) -\frac{\alpha}{4\pi}\int K(p,q;m){g}(q)\, \frac{dq}{\bar{E}_q}
,\end{align}
where $\bar{E}_p\equiv\sqrt{p^2+m_{\rm tot}^2}$ and ${g}(p)=p\bar{E}_p{f}(p) $.
Now we change to logarithmic variables $dq/\bar{E}_q = dy$ and so define $q=\sinh(y)$, $\bar{E}_q=\cosh(y)$, $p=\sinh(x)$, $\bar{E}_p=\cosh(x)$. This gives
\begin{align}
MF(x) =2\cosh(x) F(y) -  \frac{\alpha}{4\pi}\int K(x,y)\sqrt{\cosh(x)\cosh(y)}F(y)\, dy
,\end{align}
where $F(x) = {{g}(x)}/\sqrt{\cosh(x)} = p\sqrt{\bar{E}_p}{f}(p)$. When we present our form factors it is $F(x)$ that we plot. Now we have an integral equation with a completely symmetric integrand. $K(x,y)$ will correspond to either $K(p,q;m_{\rm tot})$ or $K(p,q;0)$ in which case $m$ no longer appears.

The logarithmic singularity of $K(x,y)$ at $x =y$ is integrable and so can be tamed as follows. With the cosh factors implicit we write
\begin{align}
\int K(x,y)F(y)\,dy = F(x)\int K(x,y)\, dy + \int K(x,y)[F(y)-F(x)]\,dy.
\end{align}
The first integral, containing the singularity, can be evaluated numerically to desired accuracy, while the second can now be handled through the discretization of the integral equation. We use an optimized discretization where points and weights are chosen according to the method of Gaussian quadrature. The result is a symmetric matrix eigenvalue problem that can be efficiently solved.

While we have presented results at fixed cutoff, an extrapolation to an infinite cutoff presents no difficulties away from the critical coupling. The cutoff dependent mass behaves as $M(\Lambda) \sim M_{\infty} + c\Lambda^{-2\nu}$, where $M_{\infty}$ is the infinite cutoff value and $\nu$ is the scaling dimension. It is only very close to the critical coupling where $\nu\to0$ that this extrapolation become difficult.

\begin{acknowledgments}

This research is supported in part by the Natural Sciences and Engineering Research
Council of Canada. The authors wish to acknowledge William Dykshoorn for useful conversations and Robert Reinhard for a careful reading of the manuscript.

\end{acknowledgments}	

\end{document}